\documentclass[amssymb, preprint, nobibnotes, showpacs, superscriptaddress, aps, prl]{revtex4-1}
\usepackage{graphicx,amsmath,amssymb,color}

\begin{document}

\title{Collective Multi-Photon Blockade In Cavity Quantum Electrodynamics}


\author{C. J. Zhu}
\affiliation{MOE Key Laboratory of Advanced Micro-Structured Materials, School of Physics Science and Engineering, Tongji University, Shanghai, China 200092}
\affiliation{Institute for Quantum  Science and Engineering,
	and Department of Biological and Agricultural Engineering
	Texas A\&M University, College Station, TX 77843, USA}

\author{Y. P. Yang}
\email[]{yang\_yaping@tongji.edu.cn}
\affiliation{MOE Key Laboratory of Advanced Micro-Structured Materials, School of Physics Science and Engineering, Tongji University, Shanghai, China 200092}

\author{G. S. Agarwal}
\email[]{girish.agarwal@tamu.edu}
\affiliation{Institute for Quantum  Science and Engineering,
	and Department of Biological and Agricultural Engineering
	Texas A\&M University, College Station, TX 77843, USA}
\affiliation{Department of Physics, Oklahoma State University, Stillwater, Oklahoma 74078, USA}


\date{\today}

\begin{abstract}
We present a study of collective multi-photon blockade in coherently driven atoms in a single mode cavity. Considering two atoms strongly coupled to an optical cavity, we show that the two-photon blockade with two-photon anti-bunching, and the three-photon blockade with three-photon anti-bunching can be observed simultaneously. The three-photon blockade probes both dressed states in the two photon and three photon spaces. The two photon and three photon blockades strongly depend on the location of two atoms in the strong coupling regime. The asymmetry in the atom-cavity coupling constants opens new pathways for multiphoton blockade which is also shown to be sensitive to the atomic decay and pumping strengths.
The work presented here predicts many new quantum statistical features due to the collective behavior of atoms and can be useful to generate non-classical photon pairs.
\end{abstract}


\maketitle

The well known dressed state spectrum of an atom coupled strongly to a high quality cavity has many features which have been the hall marks of the cavity quantum electrodynamics~\cite{scully,agarwal,Raimond2001}. The first doublet in the dressed states which leads to the vacuum Rabi splitting has been the subject of innumerable investigations~\cite{Mondragon1983,Agarwal1984,Raizen1989,Maunz2005,Raimond2001,Wallraff2004,Johansson2006}. An important property of the dressed state spectrum is its anharmonicity especially for low lying states. Thus the next doublet is to be studied to show the evidence of anharmonic spectrum. This can be done by studying the two photon transitions in the regime of strong atom-cavity coupling~\cite{Fink2008,Kubanek2008}. Birnbaum et. al. for the first time reported photon blockade effect in the absorption of a second photon when the photon was tuned to one of the states of the lowest doublet~\cite{Birnbaum2005}. As a result, an incident photon stream with poissonian distribution can be converted into a sub-poissonian, anti-bunching photon stream. Up to date, many works have shown that two-photon blockade phenomenon is indeed feasible for many configurations~\cite{Dayan2008}, artificial atoms on a chip~\cite{Faraon2008,Reinhard2011} and superconducting circuits~\cite{Lang2011,Hoffman2011}. In strong coupling regime, other high lying doublets in the dressed state spectrum predicts many new features of quantum nonlinear optics, which still can be explored by studying multiphoton processes~\cite{Fink2008} or by studying the dynamical behavior in strongly pumped systems~\cite{Raimond2001,Eberly1980,Brecha1999}. Photon blockade in a cavity containing a high Kerr medium has been discussed~\cite{Imamoglu1997,Rebic1999,Rebic2002}.

Recent experiments on two atoms trapped at well characterized  positions have unraveled many new aspects of the collective behavior in a high quality cavity~\cite{Neuzner2016,Reimann2015}. Interesting new results include the saturation of resonance fluorescence for constructive interference, bunching photon emission for destructive interference, and suppression of super radiant behavior due to strong back reaction of the cavity~\cite{Reimann2015}. If the atoms feel asymmetric coupling to the cavity field while these are driven symmetrically then one can obtain very large second-order photon-photon correlation $g^{(2)}$, which has been related to the predominance of two photon processes~\cite{Neuzner2016}. Since cavities allow the possibility of a wide range of parameters, the two atom system can in a different parameter domain exhibit hyper-radiance which is enhanced radiation beyond super-radiance~\cite{Pleinert2016}. These features can be understood using the possible transitions among the dressed states which depend on the two coupling constants of the cavity mode with the atoms. When the coupling constants are different then new channels open up leading to new physical effects. For example if the two coupling constants differ in phase by $\pi$, then the two atom system permits a two photon resonant process which is also one photon resonant. Thus it is worthwhile to examine the nature of not only second-order photon-photon correlation but also third-order photon-photon correlation $g^{(3)}$. The anharmonicity of the dressed state spectrum enables us to study many features of the photon blockade for a system of two atoms. 

In this work we present some interesting results for the multiphoton blockade in a collective system. Specifically we present new results when the external field is tuned to either one photon resonance or two photon resonance. We note that a recent experiment with a single atom in a high quality cavity reports higher order photon blockade~\cite{Hamsen2016} i.e., when the absorption of a third photon is forbidden if the two photon absorption is resonant. However, we find many interesting features of multiphoton blockade associated with the collective behavior of two atoms that a single atom does not possess. We show that in the case of in-phase radiation, sub-possonian statistics of third-order photon-photon correlation can be observed as a signature of three-photon blockade. We also show that the quantum statistical properties changes significantly if the atoms feels different coupling strength to the cavity. For example, in the case of $\pi/2$ phase shift, two photon blockade can be significantly improved, while three photon blockade with two photon bunching can be realized in the case of out-phase radiation. 



\begin{figure}[htb]
	\includegraphics[width=13 cm]{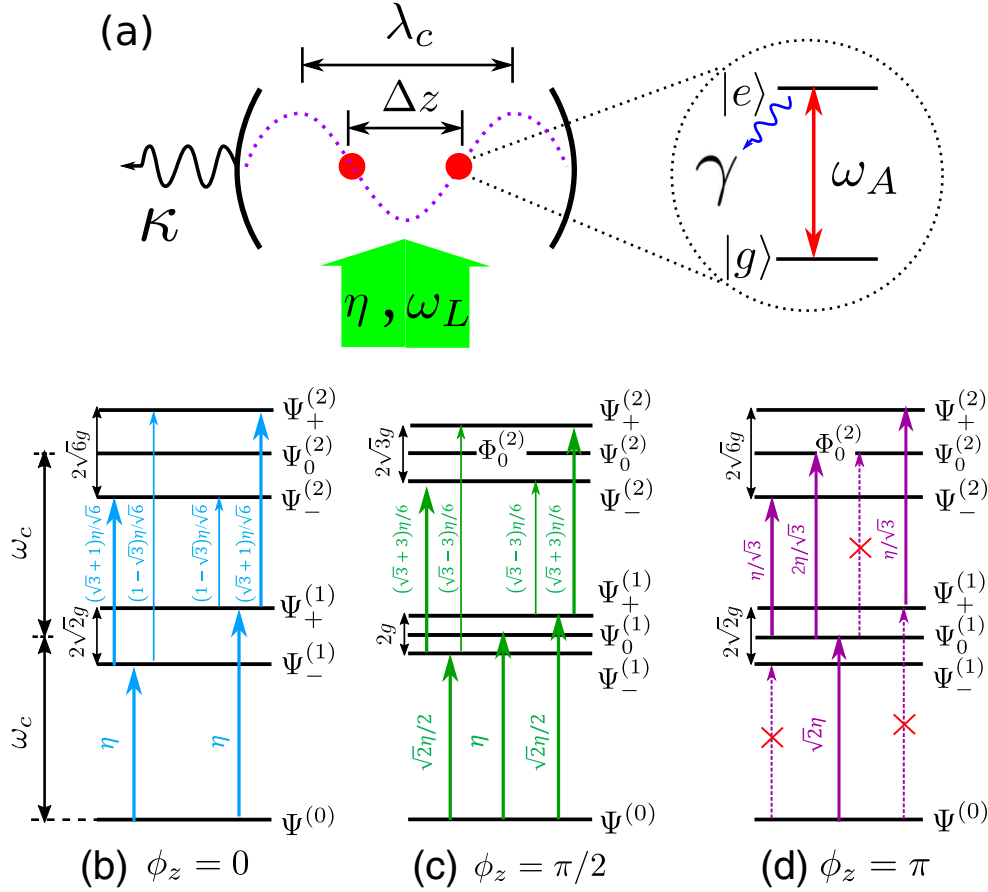}
	\caption{(a) Schematic diagram of a single mode cavity (wavelength $\lambda_c$) coupled with two two-level atoms driven by a pump field of frequency $\omega_L$. The Rabi frequency of the pump field is labeled as $\eta$. The spontaneous emission rate of the excited state for both atoms is  $2\gamma$ and the cavity decay rate is $2\kappa$. Panels (b)-(d) represent the dressed-states structure for the same coupling strength but with $\phi_z=0,\ \pi/2\ {\rm and}\ \pi$, respectively. Here, only few two photon transitions are shown in panel (c). }\label{fig:2atom}
\end{figure}

%
To realize multi-photon blockade, we now consider a scheme consisting of two two-level atoms in a single mode cavity (see Fig.~\ref{fig:2atom}(a)). 
The dynamical behavior of the system shown in Fig.~\ref{fig:2atom}(a) can also be treated according to the Jaynes-Cummings model and using master equation which is given by
\begin{equation}\label{eq:Master}
	\frac{d}{dt}\rho=-i\left[H_0+H_I+H_L,\rho\right]+{\cal L}_\gamma\rho+{\cal L}_\kappa\rho,
\end{equation}
where $\rho$ is the density matrix operator of this atom-cavity system. In a frame rotating with the frequency of the external field, the Hamiltonian of the atoms and the cavity field can be expressed as $H_0=\hbar\Delta_A(S_z^1+S_z^2)+\hbar\Delta_c (a^\dagger a)$ with $\Delta_A=\omega_A-\omega_L$ and $\Delta_c=\omega_c-\omega_L$. Here, $H_I=\hbar\sum_{i=1,2}g_i(a^\dagger S_-^i+a S_+^i)$ is the interaction Hamiltonian between atoms and cavity and the atom-cavity coupling strength is given by $g_i=g\cos{(2\pi z_i/\lambda_c)}$ which is position-dependent. $H_L=\eta\sum_{i=1,2}(S_-^i+S_+^i)$ represents the interaction Hamiltonian with the coherent pumping field. The Liouvillian ${\cal L}_\kappa\rho=\kappa(2a\rho a^\dagger-a^\dagger a\rho-\rho a^\dagger a)$ is associated with the cavity decay at rate $\kappa$, whereas ${\cal L}_\gamma\rho=\gamma\sum_{i=1,2}(2S_-^i\rho S_+^i-S_+^i S_-^i\rho-\rho S_+^i S_-^i)$ arises from the spontaneous decay of the excited state of each atom at rate $\gamma$. As in Ref.~\cite{Pleinert2016}, it is convenient to show the physical mechanism of the system by using the collective states $|gg\rangle$,  $|\pm\rangle=(|eg\rangle\pm|ge\rangle)/\sqrt{2}$ and $|ee\rangle$ as basis to rewrite the Hamiltonian. At this end, the Hamiltonian can be expressed in terms of the collective operators $D^\dagger_\pm=(S_+^1\pm S_+^2)/\sqrt{2}$, yielding $H_L=\sqrt{2}\hbar\eta(D_+^\dagger+D_+)$ and $H_I=H_++H_-$ with $H_\pm=\hbar g_\pm(aD_\pm^\dagger+a^\dagger D_\pm)/\sqrt{2}$ and  $g_\pm=g(1\pm\cos{\phi_z})$. Here, $\phi_z=2\pi\Delta z/\lambda_c$ is the phase shift between the radiation of two atoms with $\Delta z=z_2-z_1$ being the distance between two atoms.




First, we consider the simplest condition of in-phase radiation of two atoms by taking $\phi_z=0$ and $g_1=g_2=g$ (i.e., two atoms have the same coupling strength). In this case the state $|-\rangle$ is uncoupled from the interaction with cavity. Under the weak pump field approximation, it is easy to obtain a set of eigenstates $\Psi^{(1)}_{\pm}=\frac{\sqrt{2}}{2}|+,0\rangle\pm\frac{\sqrt{2}}{2}|gg,1\rangle$ in one-photon space with eigenvalues $\lambda^{(1)}_{\pm}=\hbar\omega_c\pm\sqrt{2}g\hbar$. Similarly, in two photon space, we can also obtain a set of eigenstates $\Psi^{(2)}_0=-\frac{\sqrt{3}}{3}|gg,2\rangle+\frac{\sqrt{6}}{3}|ee,0\rangle$ with eigenvalue $\lambda^{(2)}_0=2\hbar\omega_c$, and $\Psi^{(2)}_\pm=\frac{\sqrt{3}}{3}|gg,2\rangle+\frac{\sqrt{6}}{6}|ee,0\rangle\pm\frac{\sqrt{2}}{2}|+,1\rangle$ with eigenvalues  $\lambda^{(2)}_\pm=2\hbar\omega_c\pm\sqrt{6}g\hbar$ [shown in  Fig.~\ref{fig:2atom}(b)].

\begin{figure}[htb]
	\includegraphics[width=11 cm]{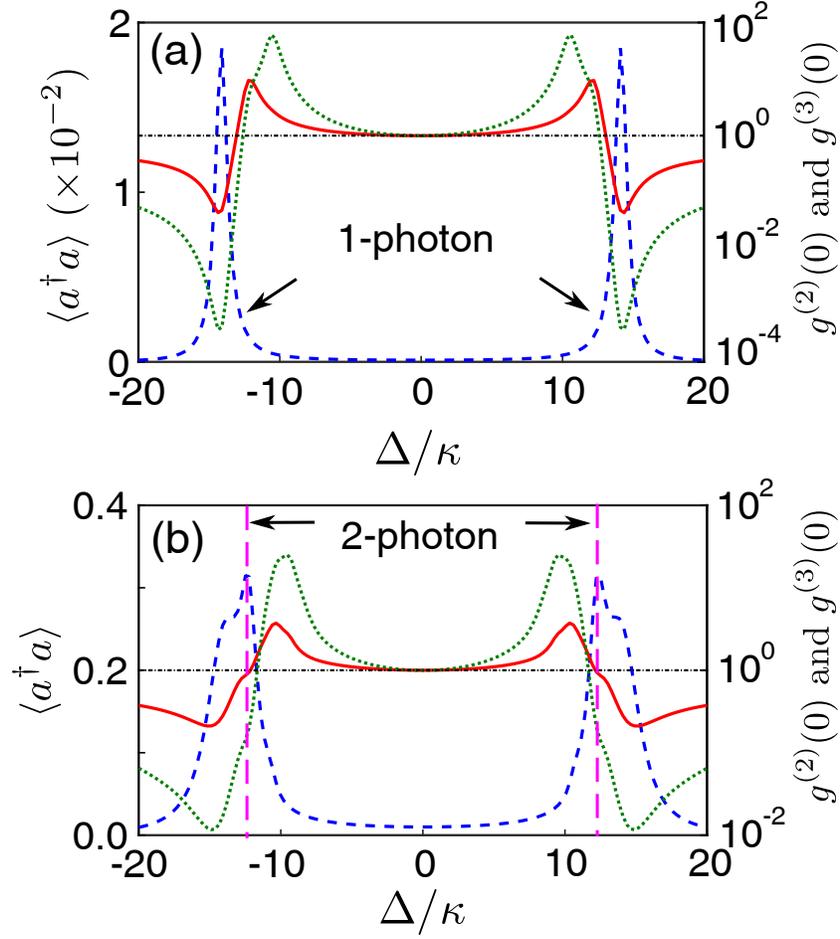}
	\caption{(Color online) The plots of field correlation function $g^{(2)}(0)$ (red solid curve), $g^{(3)}(0)$ (green dotted curve) and the mean photon number $\langle a^\dagger a\rangle$ (blue dashed curve) as a function of the normalized detuning $\Delta/\kappa$. The black dash-dotted line indicates $g^{(2)}(0)=g^{(3)}(0)=1$ for Poissonian statistics and the pink dashed lines indicate the frequencies of the two-photon excitation. Here, we choose $\eta/\kappa=0.1$ and $1$ for  panels (a) and (b), respectively. The coupling constant $g/\kappa=1$ and the decay rate $\gamma/\kappa=1$ are taken in both panels.
	}\label{fig:g2a-2a}
\end{figure}
Numerically solving Eq.~(\ref{eq:Master}) and assuming $\omega_c=\omega_A$ (i.e., $\Delta_A=\Delta_c\equiv\Delta$), we plot the mean photon number (blue dashed curve), the field correlation function $g^{(2)}(0)=\langle a^\dagger a^\dagger a a \rangle/(\langle a^\dagger a\rangle)^2$ (red solid curve) and $g^{(3)}(0)=\langle a^\dagger a^\dagger a^\dagger a a a\rangle/(\langle a^\dagger a\rangle)^3$ (green dotted curve) versus the normalized detuning $\Delta/\kappa$ in Fig.~\ref{fig:g2a-2a}. When the atoms are driven by a weak pump field (e.g., $\eta/\kappa=0.1$), we show in panel (a) that an excitation doublet can also be observed at $\Delta=\pm\sqrt{2}g$ corresponding to the one-photon excitation, leading to the transitions $\Psi^{(0)}=|gg,0\rangle\rightarrow\Psi^{(1)}_\pm$ shown in Fig.~\ref{fig:2atom}(b). Because the pump field is detuned for the transitions $\Psi^{(1)}_\pm\rightarrow\Psi^{(2)}_\pm$, the two photon transition is off resonant and thus weak leading to photon blockade. Therefore, the corresponding two-photon correlation function $g^{(2)}(0)\approx0.05<1$ is the evidence of the two-photon blockade. At these frequencies we can also find that three-photon correlation function $g^{(3)}(0)$ is extremely small. If one increases the pump field, we find that the one-photon and two-photon excitations ($\Psi^{(0)}\rightarrow\Psi^{(1)}_\pm\rightarrow\Psi^{(2)}_\pm$) are both important, resulting in four peaks in the excitation spectrum at $\Delta=\pm\sqrt{2}g$ and $\Delta=\pm\sqrt{6}g/2$, respectively (see panel (b)). Here, we take $\eta/\kappa=1$ and the pink dashed lines indicate the frequencies for the two-photon excitations. It is clear to see that the field correlation functions at $\Delta=\pm\sqrt{2}g$ satisfy $g^{(2)}(0)\approx0.3<1$ due to the two-photon blockade. The $g^{(3)}(0)$ remains much smaller than unity because the three-photon transition is highly detuned. More interestingly, we find that at $\Delta=\pm\sqrt{6}g/2$ the three-photon correlation function $g^{(3)}(0)=0.19$ accompanied with simultaneous two-photon correlation $g^{(2)}(0)$ [$\approx1.04$] nearly unity. Thus at two photon excitation condition the third photon is not absorbed leading to {\it three-photon blockade}.

%
\begin{figure}[htb]
	\centering
	\includegraphics[width=11 cm]{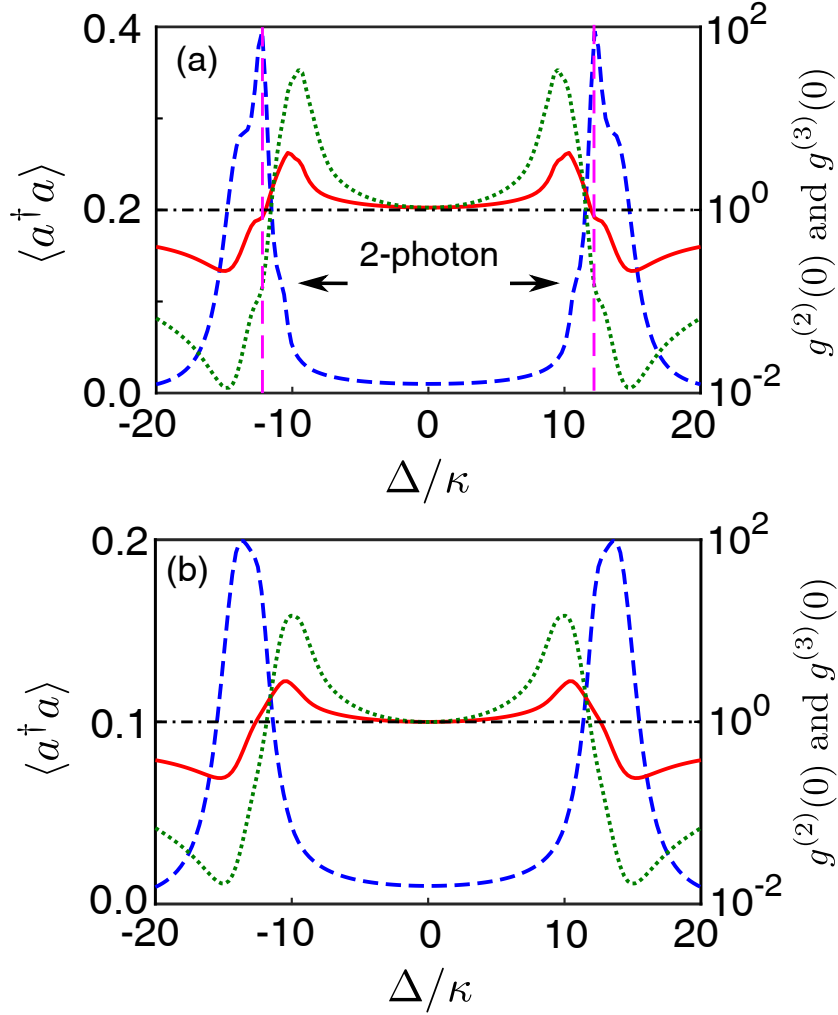}
	\caption{(Color online) Plots of the average photon number (blue dashed curve), the field correlation function $g^{(2)}(0)$ (red solid curves) and $g^{(3)}(0)$ (green dotted curves) versus the normalized detuning $\Delta/\kappa$ with $\phi_z=0$, $g/\kappa=10$ and $\eta/\kappa=1$. The black dash-dotted line indicates $g^{(2)}(0)=g^{(3)}(0)=1$ for Poissonian statistics and the pink dashed line indicates the frequency for two-photon excitation. Here, we take $\gamma=\kappa/3$ for panel (a) and $\gamma=3\kappa$ for panel (b), respectively.}\label{fig:2atom-ga}
\end{figure}
The numerical calculations show that the blockade is sensitive to the atomic decay rate $\gamma$ as compared to $\kappa$. This is illustrated in Fig.~\ref{fig:2atom-ga}. For a small atomic decay rate, e.g. $\gamma=\kappa/3$, the two-photon transition can be excited much more easily than in case of $\gamma=\kappa$, so that the mean photon number at $\Delta=\pm\sqrt{6}g/2$ becomes much larger than that at $\Delta=\pm\sqrt{2}g$ and the three-photon blockade can also be observed with $g^{(2)}(0)\approx0.8$ [see Fig.~\ref{fig:2atom-ga}(a)]~\cite{note1}. However, In the case of large decay rate, i.e. $\gamma=3\kappa$, the two-photon transition can hardly be excited since the atom rapidly decays to the ground state after one-photon transition takes place. Therefore, the mean photon number decreases slightly and the peaks associated with the one-photon and two-photon excitations merge together [shown in Fig.~\ref{fig:2atom-ga}(b)]. As a result, the width of the peak becomes large. It is worth to point out that at $\Delta=\pm\sqrt{6}g/2$ we can obtain $g^{(2)}(0)\approx1.25$, $g^{(3)}(0)\approx0.4$, and we can find $g^{(2)}(0)\approx0.32$, $g^{(3)}(0)\approx0.02$ at $\Delta=\pm\sqrt{2}g$, indicating that we can realize two-photon and three-photon blockade simultaneously. 

In following, we study the case of two atoms for $\phi_z\neq0$ so that $g_1\neq g_2$. In this case additional pathways open up as the asymmetric state $|-\rangle=(|eg\rangle-|ge\rangle)/\sqrt{2}$ also participates in the transitions. To understand the physical mechanism, one can use a set of collective states to describe the atomic transitions. Defining $\alpha=(1+\cos\phi_z)/\sqrt{1+\cos^2\phi_z}$ and $\beta=(-1+\cos\phi_z)/\sqrt{1+\cos^2\phi_z}$, we can find three eigenstates in one-photon space, i.e.,  $\Psi^{(1)}_0=\frac{\sqrt{2}}{2}\alpha|-,0\rangle+\frac{\sqrt{2}}{2}\beta|+,0\rangle$ with eigenvalue $\lambda^{(1)}_0=\hbar\omega_c$ and $\Psi^{(1)}_\pm=\mp\frac{\sqrt{2}}{2}|gg,1\rangle-\frac{1}{2}\beta|-,0\rangle+\frac{1}{2}\alpha|+,0\rangle$ with eigenvalues $\lambda^{(1)}_\pm=\hbar\omega_c\pm\hbar g_0\sqrt{1+\cos^2{\phi_z}}$, respectively. In two photon space, we can obtain four eigenstates, labeled as  $\Psi^{(2)}_0=-\frac{\sqrt{3}}{3}|gg,2\rangle+\frac{\sqrt{6}}{3}|ee,0\rangle$ and  $\Phi^{(2)}_0=\frac{\sqrt{2}}{2}\alpha|-,1\rangle+\frac{\sqrt{2}}{2}\beta|+,1\rangle$ with the same eigenvalue $\lambda^{(2)}_0=2\hbar\omega_c$ and $\Psi^{(2)}_\pm=\frac{\sqrt{3}}{3}|gg,2\rangle\mp\frac{1}{2}\beta|-,1\rangle\pm\frac{1}{2}\alpha|+,1\rangle+\frac{\sqrt{6}}{6}|ee,0\rangle$ with eigenvalues $\lambda^{(2)}_\pm=2\hbar\omega_c\pm\hbar g\sqrt{3(1+\cos^2{\phi_z})}$, respectively. As shown in Fig.~\ref{fig:2atom}(b)-(d), one can also obtain the transition strength by calculating the dipole matrix element of operator  $\sqrt{2}\eta(D_+^\dagger+D_+)$. These dressed states along with some of the important transitions are shown in Fig.~\ref{fig:2atom}(c) and (d).

\begin{figure}[htb]
	\centering
	\includegraphics[width=15 cm]{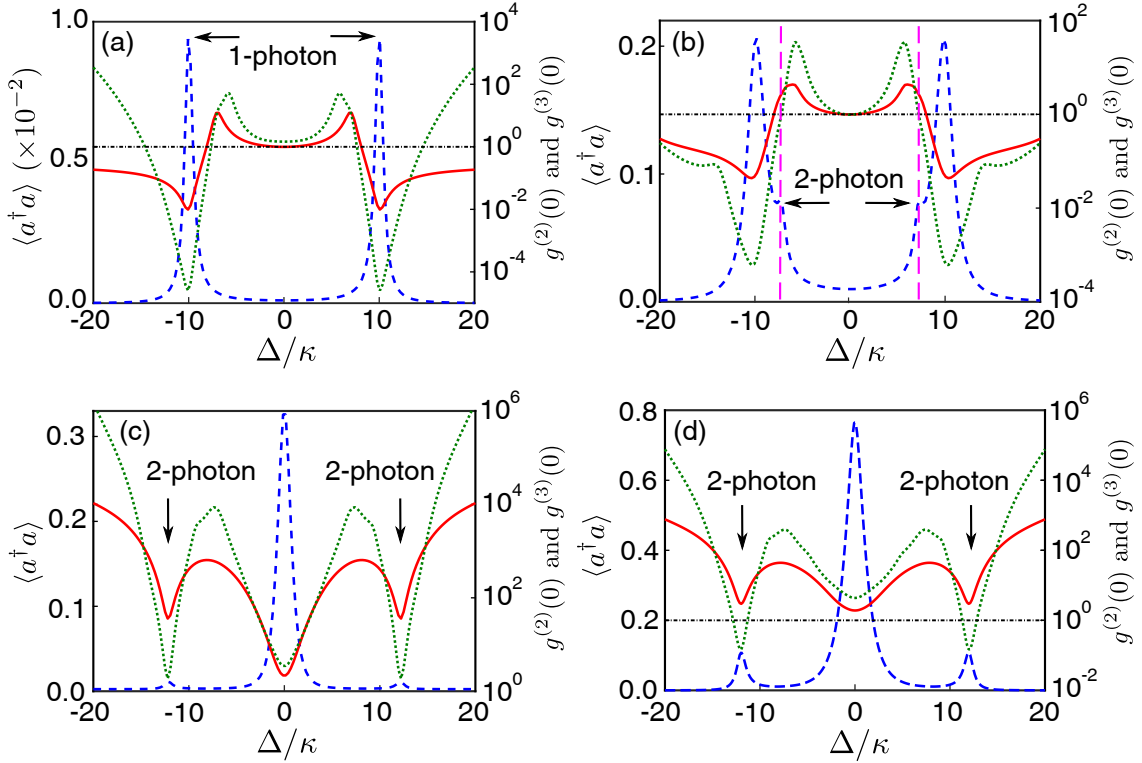}
	\caption{(Color online) Plots of the average photon number (blue dashed curve), the field correlation function $g^{(2)}(0)$ (red solid curves) and $g^{(3)}(0)$ (green dotted curves) versus the normalized detuning $\Delta/\kappa$. The phase shift is chosen as $\phi_z=\pi/2$ (panels (a,b)) and $\pi$ (panels (c,d)), respectively. The pump field $\eta/\kappa=0.1$ (a), $1.0$ (b,c) and $2.0$ (d), respectively.}\label{fig:2atom-phi}
\end{figure}
In Fig.~\ref{fig:2atom-phi}, we plot the mean photon number $\langle a^\dag a\rangle$ and field correlation functions $g^{(2)}(0)$ and $g^{(3)}(0)$ versus the normalized detuning $\Delta/\kappa$ with different pump field and phase shift. We show that the properties of the quantum statistics of the system changes significantly with the collective behavior of two atoms. First, we consider the case of $\pi/2$ phase shift with pump field $\eta=0.1\kappa$ (a) and $\eta=\kappa$ (b), we find that the one-photon excitations are dominant although the eigenstate become much more complicated as shown in Fig.~\ref{fig:2atom}(c). Now the one photon excitations occur at  $\Delta=\pm g$. Comparing the results of in-phase radiation (see Fig.~\ref{fig:g2a-2a}), it is obvious that two-photon blockade phenomenon is even more pronounced. We find that $g^{(2)}(0)\approx0.01$ in panel (a) and $g^{(2)}(0)\approx0.03$ in panel (b), which is nearly $10$ times less than that of the in-phase condition for a strong pump field. This characteristic can be explained by calculating the dipole matrix elements associated with the one-photon and two-photon excitations. In $\pi/2$ case, the two-photon transition strength is much smaller than that in in-phase condition, so that the second-order photon-photon correlation drops significantly. For the out-phase case (i.e., $\phi_z=\pi$), the two-photon excitation becomes important because $\Psi^{(0)}\rightarrow\Psi^{(1)}_\pm$ transitions are not allowed (see Fig.~\ref{fig:2atom}(d)). As a result, we can observe a single peak at the excitation spectrum and simultaneously both field correlation functions $g^{(2)}(0)$ and $g^{(3)}(0)$ are larger than unity because the two-photon and the three-photon pathways are fully resonant [see panels (c) and (d) in Fig.~\ref{fig:2atom-phi}]. If we increase the strength of the pump $\eta=2\kappa$ [Fig.~\ref{fig:2atom-phi}(d)], then the two-photon excitations $\Psi^{(0)}\rightarrow\Psi^{(2)}_\pm$ become strong enough to be observed. In the excitation spectrum shown in Fig.~\ref{fig:2atom-phi}(c), there exist three peaks associated with the one-photon excitations ($\Delta=0$) and two-photon excitations ($\Delta=\pm\sqrt{6}g/2$) respectively. At $\Delta=0$, we  find $g^{(2)}(0)\approx2.15$ and $g^{(3)}(0)\approx3.45$ since the two photon and three photon pathways are fully resonant. At $\Delta=\pm\sqrt{6}g/2$, we find $g^{(2)}(0)\approx36.55$ and $g^{(3)}(0)\approx1.89$ so that the three-photon blockade is not well developed for very weak pump. Increasing the pump field, the three-photon blockade occurs with two-photon bunching as shown in panel (d). The corresponding field correlation functions are given by $g^{(2)}(0)\approx3.54$ and $g^{(3)}(0)\approx0.18$. We  have thus seen that the collective two atom system can lead to a wide variety of the multiphoton blockade.

\begin{figure}
	\includegraphics[width=11 cm]{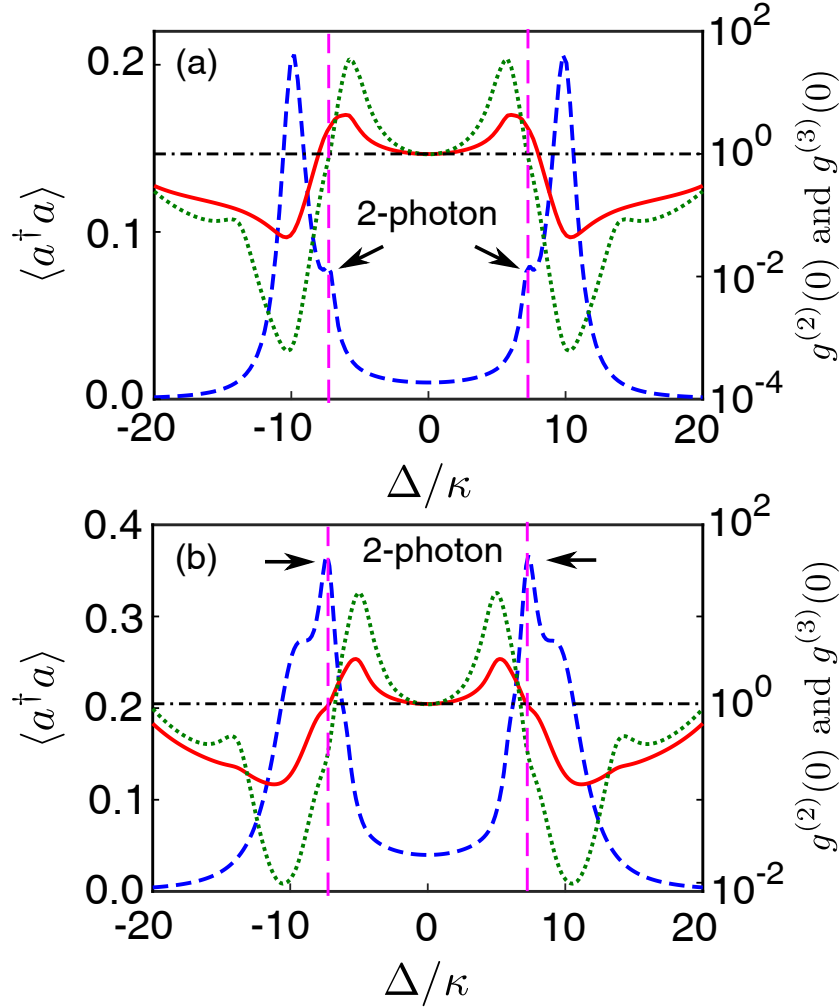}
	\caption{(Color online) The plots of the field correlation function $g^{(2)}(0)$ (red solid curve), $g^{(3)}(0)$ (green dotted curve) and the mean photon number (blue dashed curve) as a function of the normalized detuning $\Delta/\kappa$. The black dash-dotted line indicates $g^{(2)}(0)=g^{(3)}(0)=1$ for Poissonian statistics and the pink dashed line indicates the frequency for two-photon excitation. Here, we choose $\eta/\kappa=1$ for panel (a) and $2$ for panel (b). }\label{fig:g2a}
\end{figure}
We conclude this paper by comparing results with the single atom case [see Fig.~\ref{fig:g2a}]. 
In panel (a), we take $\eta/\kappa=1.0$, and it is clear to see that there exist four peaks in the excitation spectrum corresponding to the detuning $\Delta=\pm g$ and $\Delta=\pm\sqrt{2} g/2$, respectively. At $\Delta=\pm g$, two-photon blockade phenomenon can be observed since the two-photon excitations are non-resonance. This has been widely studied in literature~\cite{Birnbaum2005,Faraon2008,Reinhard2011}. The small peaks at $\Delta=\pm \sqrt{2}g/2$ correspond to the two-photon excitation process i.e., the $\Psi^{(0)}\rightarrow\Psi^{(1)}_\pm\rightarrow\Psi^{(2)}_\pm$ transitions if the pump field frequency satisfies $\omega_L=\omega_c\pm \sqrt{2}g/2$. As shown in Fig.~\ref{fig:g2a}(a), the photon-photon correlation satisfy $g^{(2)}(0)\approx3.04$ and $g^{(3)}(0)\approx1.42$ at $\Delta=\pm \sqrt{2}g/2$. When increasing the pump field, i.e., $\eta/\kappa=2$, we find that the third-order photon-photon correlation $g^{(3)}(0)\approx0.42$ and the second-order photon-photon correlation $g^{(2)}(0)\approx1.07$ [see panel (b)]. Thus the allowed two photon transition leads to $g^{(2)}(0)>1$, however the three photon blockade occurs. The three photon blockade has been recently reported~\cite{Hamsen2016}.


In conclusion, we have shown that a collective system of two two level atoms in a high quality cavity can display a much richer variety of photon blockade phenomena. This arises from the dressed state structure of the system of interacting cavity field and atoms. The dressed state structure and allowed transitions depend on the two coupling constants of the atoms to the cavity field.  We have found that in the case of in-phase radiations of two atoms, the three-photon blockade can be observed with two-photon bunching if the excitation frequency is tuned to two photon transition.  The collective behavior depending on the relative phase i.e., location of atoms can change significantly the properties of quantum statistics of the system. In case of $\pi/2$ phase shift two-photon blockade can become quite prominent even under the condition of a weak pump field, while the photon blockade phenomenon disappears if the coupling constants $g_1$ and $g_2$ are out of phase. Thus the photon blockade effects in two atom system can be quite different from the case of single-atom. The two-atom system displays very pronounced three-photon blockade which should be observable in experiments of the type reported recently. The three photon blockade should be useful in generating non-classical photon pairs.

%


\vskip 5pt

\begin{acknowledgments}
	Dr. C. J. Zhu and Y. P. Yang acknowledge the National Key Basic Research Special Foundation (Grant No. 2016YFA0302800); the 973 Program (Grant No. 2013CB632701); the Joint Fund of the National Natural Science Foundation (Grant No. U1330203); the National Nature Science Foundation (GrantNo. 11504272); the Shanghai Science and Technology Committee (STCSM) (Grants No. 15XD1503700). C. J. Zhu also thanks Marc Pleinert for advice on the numerical code. G. S. Agarwal thanks the Biophotonics initiative of the Texas A\&M University for support.
\end{acknowledgments}


\begin{thebibliography}{99}
\bibitem{note1} Decreasing the pump field, one can also realize three-photon blockade with two-photon bunching in case of $\gamma=\kappa/3$.
\end{thebibliography}
\end{document}